\documentclass[a4paper, conference]{IEEEtran}
\usepackage{cite}
\usepackage{amsmath,amssymb,amsfonts}
\usepackage{algorithmic}
\usepackage{graphicx}
\usepackage{textcomp}
\usepackage{xcolor}
\usepackage{balance}

\begin{document}
\IEEEoverridecommandlockouts
\IEEEpubid{\makebox[\columnwidth]{979-8-3503-2400-6/23/\$31.00~\copyright 2023 IEEE \hfill}\hspace{\columnsep}\makebox[\columnwidth]{ }}

\title{FDTD Full Wave Simulations of Reconfigurable Intelligent Surfaces
}

\author{
E. Colella$^{1,2}$, L. Bastianelli$^{1,2}$, V. Mariani Primiani$^{1,2}$, F. Moglie$^{1,2}$ \\
  1 CNIT, Viale G.P. Usberti 181/A, 43124, Parma, Italy\\
  2 DII, Università Politecnica delle Marche, via Brecce Bianche 12, 60131, Ancona, Italy\\

}


\maketitle

\begin{abstract}
This paper presents the analysis of metasurfaces, here called reconfigurable intelligent surface. The analysis is performed by numerical simulations that implement the finite-difference time-domain method. The metasurface has been modeled by metallic patches interconnected by varactor diodes. The electromagnetic source consists of randomly generated plane wave. This kind of analysis allows us to investigate the response of the metasurface when it is hit by a random source. 

\end{abstract}


\section{Introduction}

Metamaterials are artificially created materials, obtained thanks to the periodic arrangement of dielectric and metallic elements with sub-wavelength dimensions and spacings much smaller than the wavelength \cite{4}.
These materials display unique electromagnetic properties and have attracted growing interest in science and technology over the past fifteen years \cite{5,1,2,urbas,tretyakov}.
Thanks to metamaterials it is possible to obtain unconventional light-matter interaction effects, such as a negative refractive index or super lenses.
Furthermore, thanks to the correct positioning of the elements constituting the metamaterial, it is possible to manipulate the electromagnetic fields. However, practical 3D applications of metamaterials have been hampered by significant structural challenges \cite{16, 17}. For this reason, interest has focused more on 2D implementations, called metasurfaces, which are much simpler to implement, less expensive and more manageable, being compatible with already established planar machining techniques on silicon. Electromagnetic metasurfaces are based on the idea of radio frequency transmitting/receiving antenna arrays, consisting of an array of 2D resonant elements capable of locally modifying both the phase and the amplitude of an incident electromagnetic wave \cite{18, 19, 20}. Furthermore, thanks to the use of plasmon or high refractive index dielectric effects, the dimensions of the metasurfaces can be much smaller than the wavelength. Starting from the work developed by Yu et al. based on nano-antennas, which is capable of modifying the wavefront of an electromagnetic wave to obtain anomalous reflection/refraction effects \cite{7}. This has led to the creation of various models to increase the bandwidth and efficiency of these devices, opening the door to a new research sector based on 2D surface optics and photonics. Furthermore, it has been demonstrated that it is possible to use reconfigurable dielectric or semiconductor resonators to realize smart screens able to control the propagation of electromagnetic fields in telecommunications. This type of screens with the possibility of automatically reconfiguring themselves according to the environmental radio conditions are called reconfigurable intelligent surfaces RISs \cite{130, emilio1, emilio2}. Reconfigurable intelligent surfaces (RIS) are surfaces made up of antennas or radiating elements that can be dynamically modified to control and manipulate the propagation of electromagnetic waves in a wide range of frequencies \cite{140}. RIS can improve the performance of wireless communication, reduce interference, extend coverage, increase the security of wireless networks and improve the energy efficiency of communication devices \cite{150}. RIS represents a promising technology for future wireless communication networks and can be used in various scenarios, presenting a great opportunity to enhance 5G communications in future smart radio environments. Within this paper, we evaluated the performance of RISs in different electromagnetic conditions by computational electrodynamics simulations to analyze their impact on the smart radio environment.

\section{Simulation set-up}
\begin{figure}[!h]
  \centering
  \includegraphics[width=70mm]{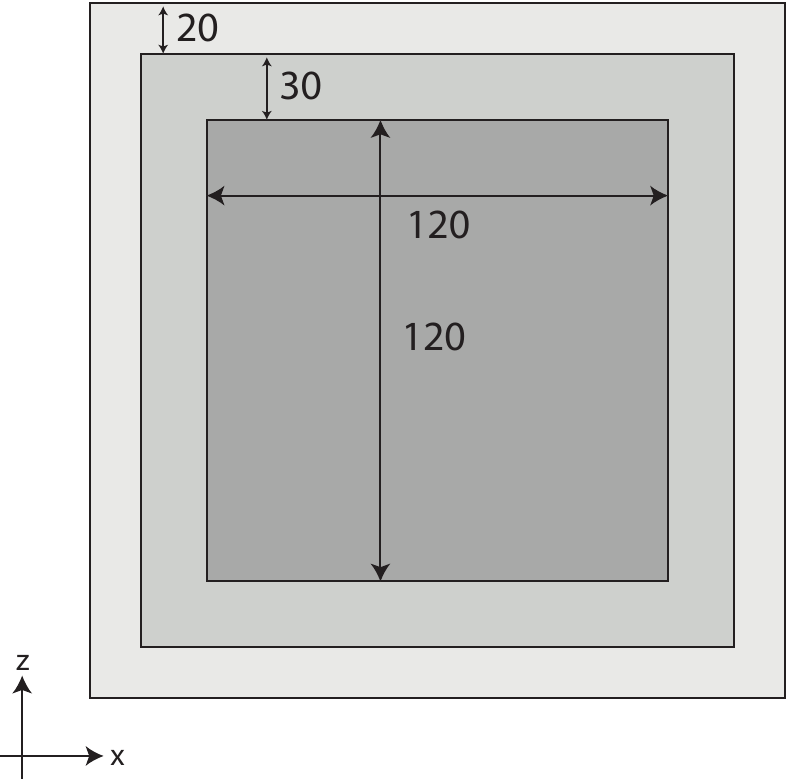}
  \caption{FDTD domains. The innermost domain is the RIS domain, the domain in the middle is the total field domain while the outermost domain is the scattered domain.}
  \label{fig:domains}
\end{figure}

Electromagnetic simulations involve analyzing the dynamic re-configurations of a RIS to evaluate its focusing performance.
To study the electromagnetic response of the RIS, numerical simulations were carried out using the finite difference time domain (FDTD) method~\cite{taflove}.
The FDTD method is a numerical technique used to reproduce the behavior of electromagnetic waves using a full wave approach.
This technique allows to study complex geometries of different kinds with high precision, allowing to obtain very accurate results~\cite{moglie2,RC_2016_Moglie_Perdite}.
The entire FDTD code has been implemented following the standard mathematical procedure.
Our group developed a parallel home made C code which run on supercomputer.
The simulations have been performed in a working space of $220 \times 110 \times 220$ for a total of $5.52 \cdot 10^{6}$ cubic cells.
The cell size is 1 mm. The electromagnetic source that hits the RIS is a plane wave randomly generated.
The generation of plane wave(s) is done by a dedicated script and then read by the FDTD code.
The whole simulated domain is divided in three domains: i) RIS domain of $120 \times 10 \times 120$ cells; ii) the total field domain of $30 \times 30 \times 30$ cells; iii) the scattered field domain of $20 \times 20 \times 20$ cells, as represented in the \figurename~\ref{fig:domains}.
A detailed explanation of the plane wave generation and the implementation of separation plane between the total field and the
scattered field region are reported in~\cite{Pastore}.
To analyze the electromagnetic behavior of metasurfaces in 5G radio environment the frequency band of 0.8 GHz -- 8.4 GHz with $f_0$=4.6 GHz has been chosen.
The angle of incidence in spherical coordinates are: $\alpha = 1.57$, $\theta = 0.78$ and $\phi=0.78$ for all simulations~\cite{Pastore}. 
\begin{figure}[!h]
  \centering
  \includegraphics[width=70mm]{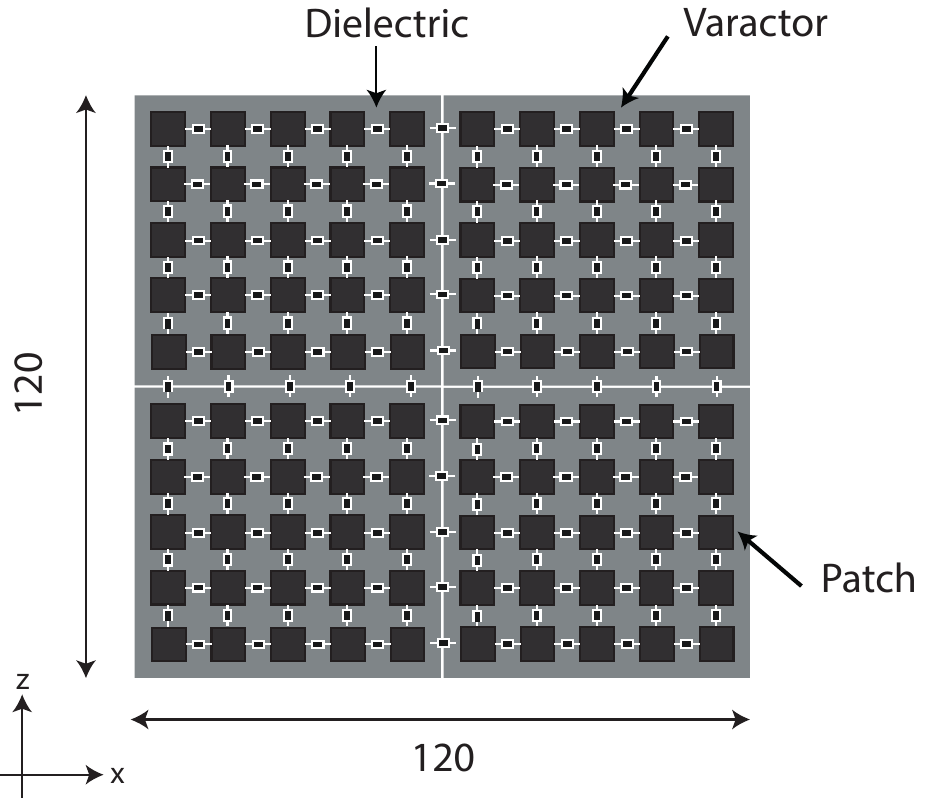}
  \caption{RIS configuration. Representation of the simulated RIS of $10 \times 10$ PEC patches of 1 cm$^2$ connected by 180 varactor diodes, vertically and horizontally polarized. The dielectric support of $\epsilon_r = 3$ and 1 mm of thickness.}
  \label{fig:ris}
\end{figure}
\begin{figure}[!h]
  \centering
  \includegraphics[width=75mm]{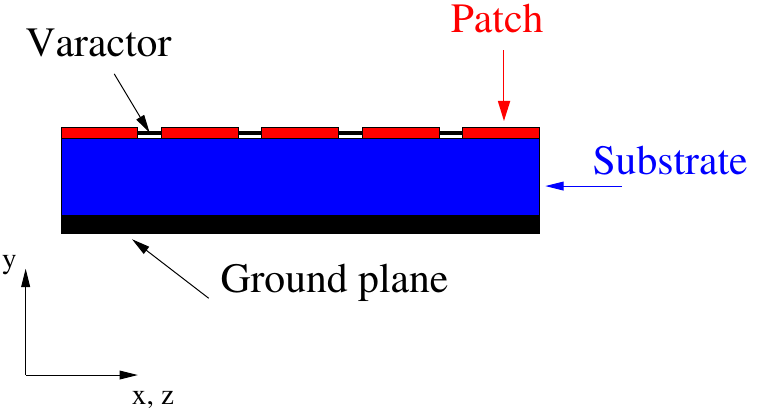}
  \caption{Side view of the simulated RIS. On the top are reported the PEC patches interconnected by varactor diodes. The substrate layer is in the middle whereas on the bottom is reported the ground plane.}
  \label{fig:geometry}
\end{figure}
The whole RIS is composed by~\cite{costa}: i) metal patches; ii) varactor diodes; iii) dielectric substrate; iv) ground plane. Figure~\ref{fig:geometry} reports the side view of the simulated RIS. The substrate, i.e. the dielectric support, has been simulated with $\epsilon_r = 4.4$ and $\sigma=0.0025$~S/m whereas the ground plane is simulated as perfect electric conductor (PEC).

\begin{figure}[!h]
  \centering
  \includegraphics[width=75mm]{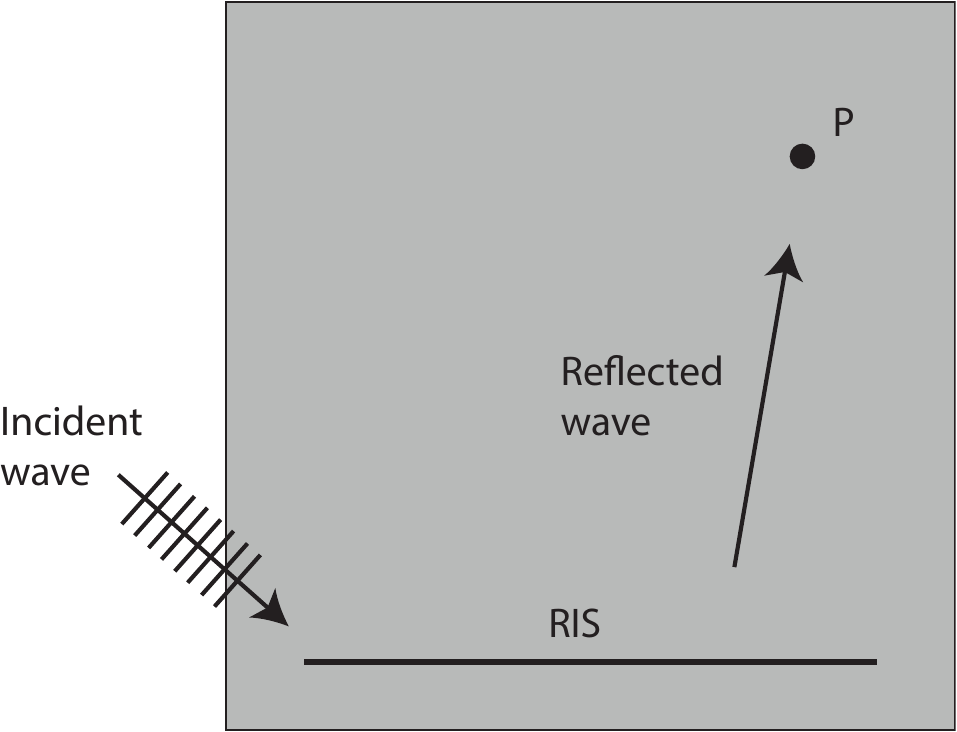}
  \caption{Simulation set-up. Representation of the simulation configuration with RIS, plane wave and detecting point P.}
  \label{fig:sim_setup}
\end{figure}
RIS consists of $10 \times 10$ resonant structures, i.e. square metal patches modeled by PEC.
Each patch is interconnected each other by 180 varactor diodes. The varactor diodes are designed as 1 mm$^3$ cell connecting two patches as shown in the \figurename~\ref{fig:ris}, while a single patch communicates with two other patches. The capacitance of the varactor is simulated simply by forcing the corresponding constitutive relationship for the fields in the FDTD cell~\cite{taflove}. The RIS is placed in the plane $xy$. The dielectric material inside the working domain was air and the absorbing boundary conditions for the whole domain are the perfect matched layers (PMLs). The values of the diodes have been 1 pF and 0.1 pF simulating a short circuit or an open circuit. The electric and magnetic fields have been recorded in different points $P$ as shown in the \figurename~\ref{fig:sim_setup}. The temporal step was 1.5 ps for a total of 100 periods in all simulations. In all simulations the RIS was placed on the $xz$ plane at the center of the workspace. The simulation set-ups are shown in the \figurename~\ref{fig:sim_setup}
During the analysis we also simulated the RIS as a PEC surface.

\section{Results}
\begin{figure}[!h]
  \centering
  \includegraphics[width=65mm]{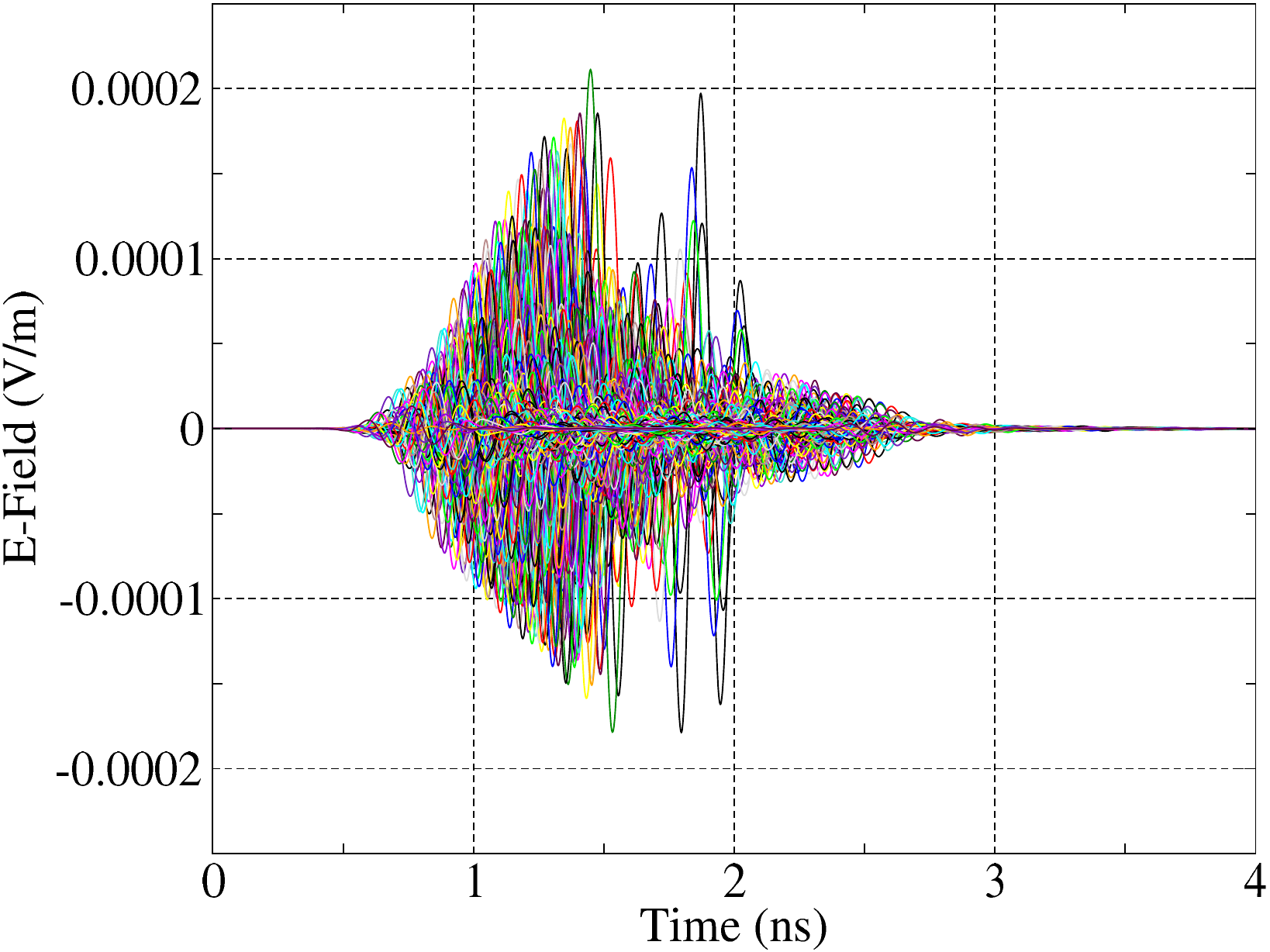}
  \caption{First simulation results. Electric field on the plane of $50 \times 50 $ cells at high of 50 cells without RIS.}
  \label{fig:result1}
\end{figure}

\begin{figure}[!h]
  \centering
  \includegraphics[width=70mm]{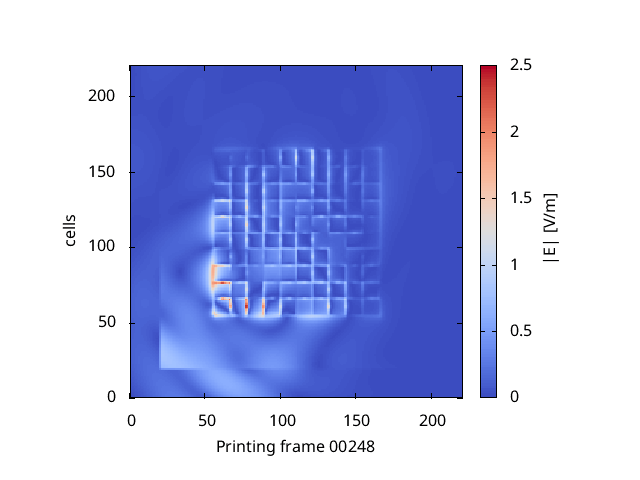}
  \caption{Field distribution of a 2D plane of the 3D domain. All the capacitor are set to 1.0 pF.}
  \label{fig:anim2}
\end{figure}

In the first simulation, to test the effectiveness of the separation between the near field and the far field, we reported the electric field on a floor consisting of $10 \times 10$ points in the case without RIS at a height of 50 cells.
The result is shown in the \figurename~\ref{fig:result1}.
The reflected field values obtained are lower than 0.2 mV/m.
 \begin{figure}[!h]
  \centering
  \includegraphics[width=65mm]{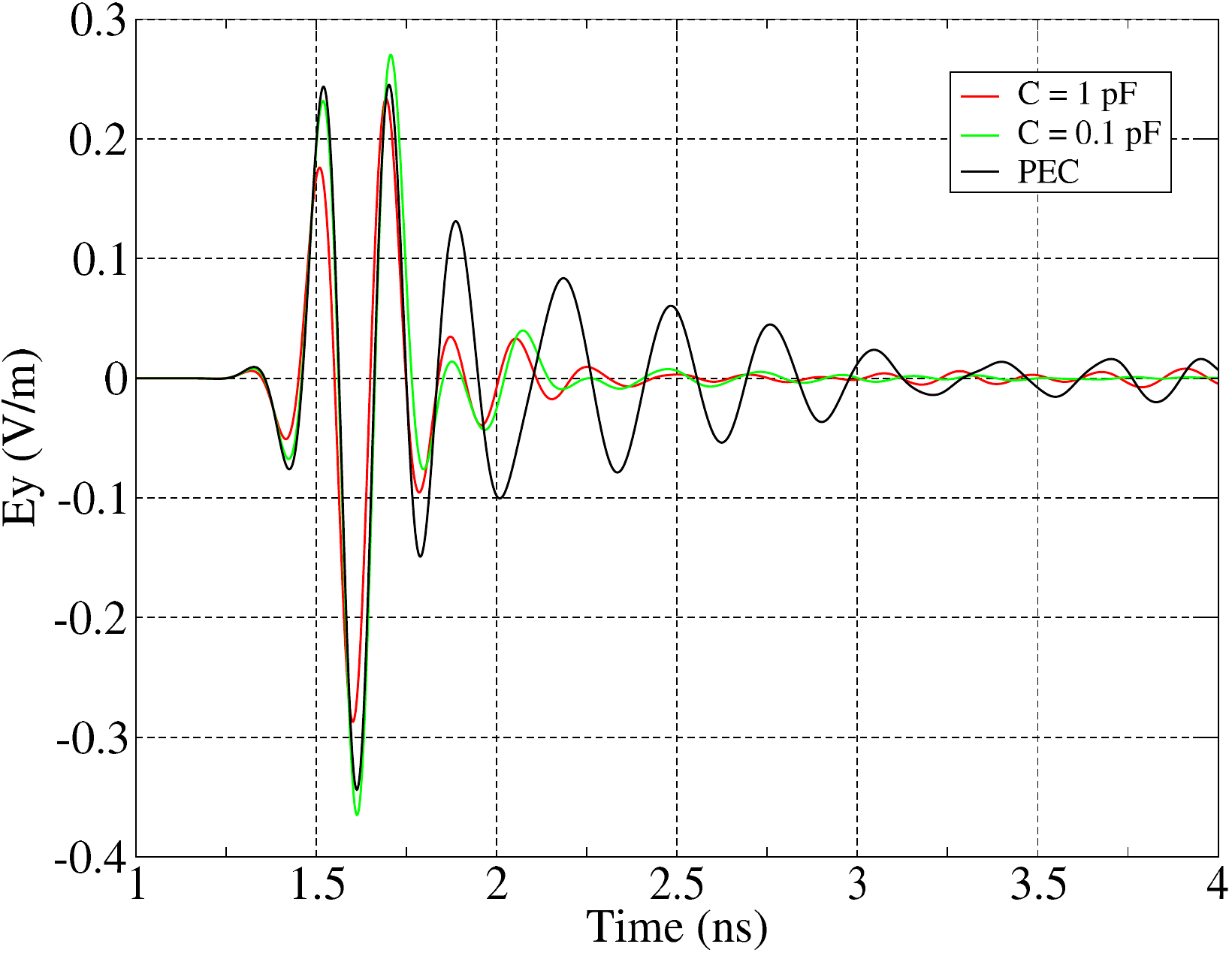} 
  \caption{Second simulation results. Electric field $P(10, 100, 10)$ in the case of  PEC surface and in the case of RIS with diodes at 1 pF and 0.1 pF configuration.}
  \label{fig:result2}
\end{figure}
 \begin{figure}[!h]
  \centering
  \includegraphics[width=65mm]{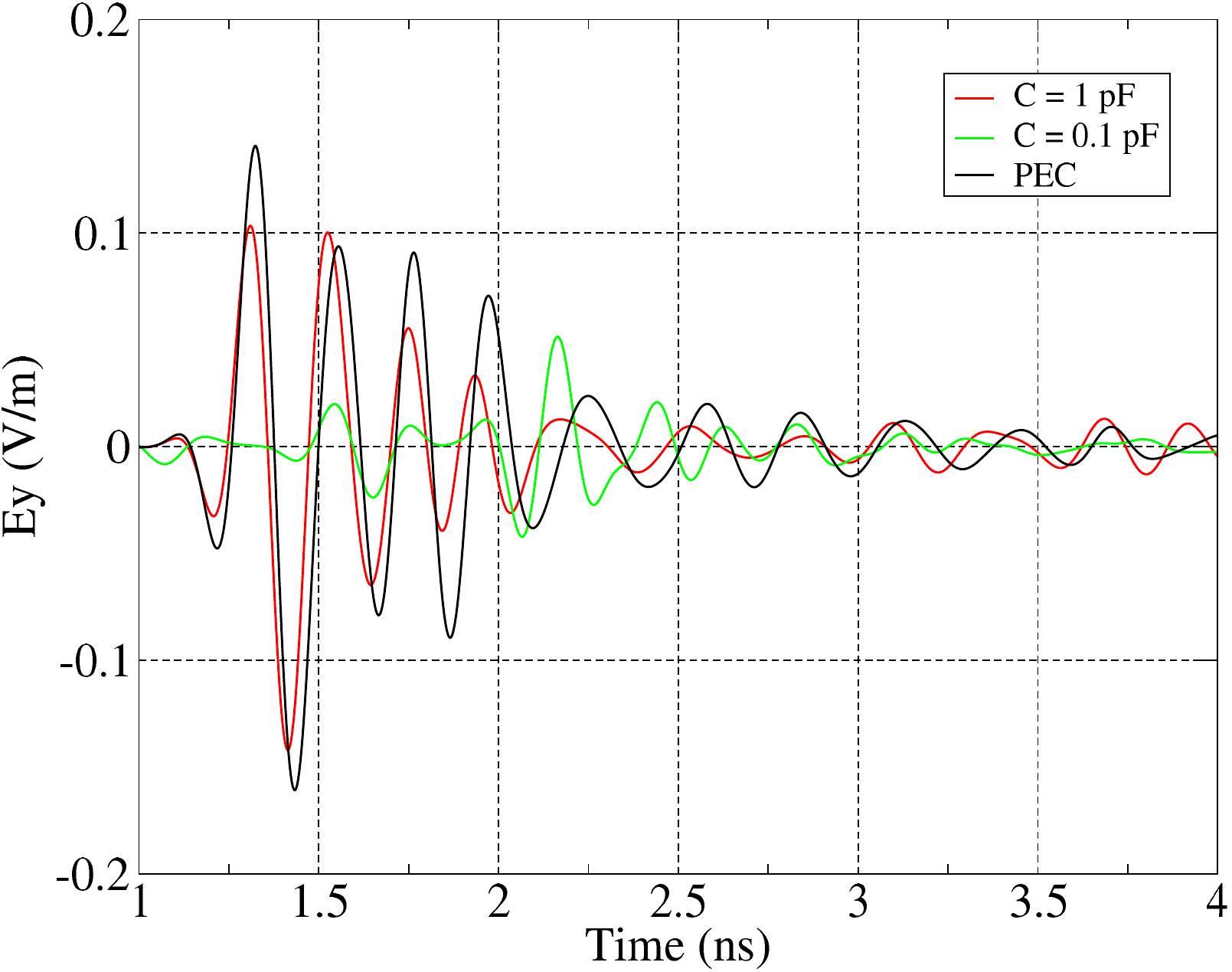}
  \caption{Third simulation results. Electric field $P(10, 100, 100)$ in the case of  PEC surface and in the case of RIS with diodes at 1 pF and 0.1 pF configuration.}
  \label{fig:result3}
\end{figure}
In that plot are reported all the probed points, namely $100$~points where the three Cartesian components of the electric field $(E_x, E_y$ and $E_z)$ are collected.
In the second case, we reported the electric field value, only the $E_y$ component, collected at the point $P(10, 100, 10)$. In this case
we compared the electric field when the capacitors of the RIS are set to $1.0$~pF and~$0.1$~pF respectively. Moreover, in the same plot, the two capacitor configurations
are compared to a RIS simulated by a PEC surface, \figurename~\ref{fig:result2}.
 \begin{figure}[!h]
  \centering
  \includegraphics[width=75mm]{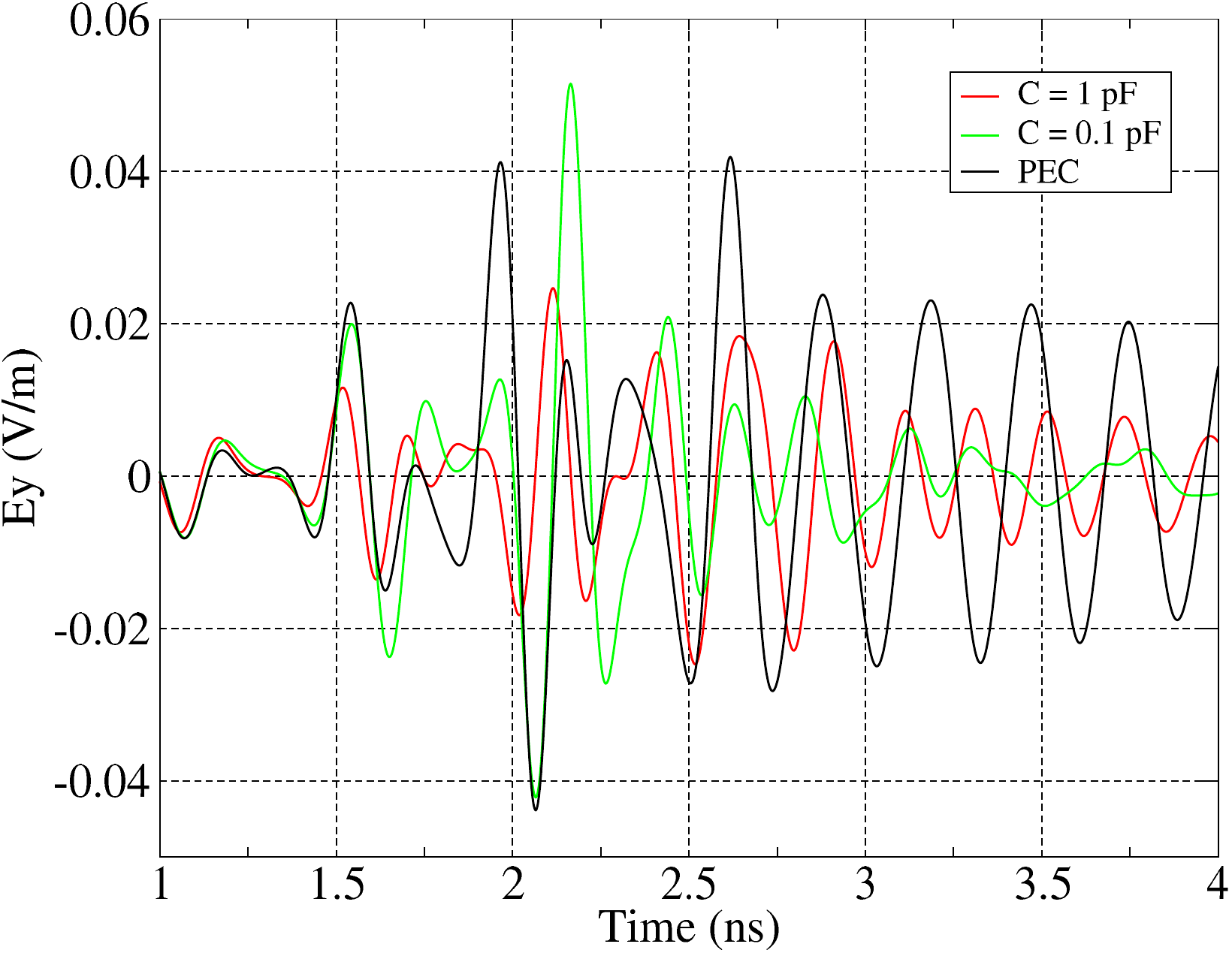} 
  \caption{Fourth simulation results. Electric field $P(210, 100, 210)$ in the case of  PEC surface and in the case of RIS with diodes at 1 pF and 0.1 pF configuration.}
  \label{fig:result4}
\end{figure}
\begin{figure}[!h]
  \centering
  \includegraphics[width=80mm]{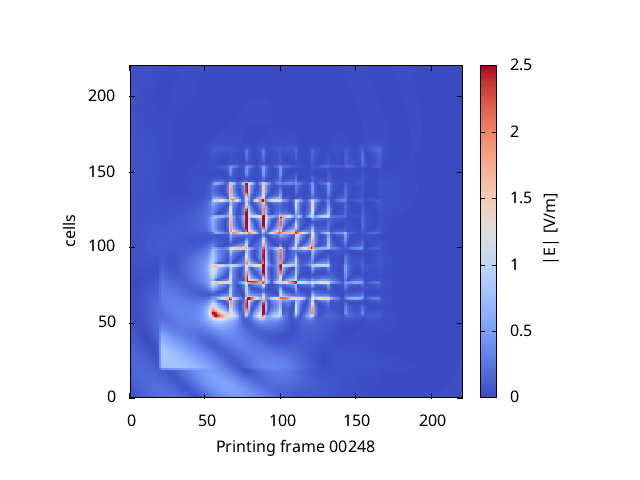}
  \caption{Field distribution of a 2D plane of the 3D domain. All the capacitor are set to 0.1 pF.}
  \label{fig:anim1}
\end{figure}
In this point ($P(10, 100, 10)$) there are no difference between the diodes configurations.
We did the same comparison in terms of electromagnetic response of the previous configurations at the point $P(10, 10, 100)$, as shown in the \figurename~\ref{fig:result3}.
We also investigate the electric field $(E_y)$ at point $P(210, 10, 210)$ with respect to the two different capacitor configurations and by the RIS simulated as PEC surface,~\figurename~\ref{fig:result4}.
In Figs.~\ref{fig:result3} and~\ref{fig:result4} the electric field $(E_y)$ has some differences based on capacitor configurations while a higher value is with the PEC surface. In Figs.~\ref{fig:result2},~\ref{fig:result3} and~\ref{fig:result4} the main time oscillation corresponds to the central frequency, $4.6$~GHz, of the spectrum of the exciting pulse.
Figure~\ref{fig:anim1} and~\figurename~\ref{fig:anim2} report the field distributions for a 2D plane of the 3D simulated domain.
The field distribution is evaluated over the RIS surface. In~\figurename~\ref{fig:anim1} all capacities are set to $0.1$~pF whereas in \figurename~\ref{fig:anim2} all
of them are set to $1.0$~pF.
In the first configuration, the module of the electric field distribution over the RIS is higher with respect to the case when all capacities are set to $1.0$~pF.

\section{Discussion and Conclusion}

In this work we reported the preliminary analysis of RIS performed by FDTD simulations. The electromagnetic source that hits the RIS consists of a random
generated plane wave. We investigated the capability of RIS to reflect and focus the electromagnetic field in a fixed spatial point within the simulation domain,
in particular inside the scattered field domain.
Currently, two RIS configurations have been simulated, in particular when all capacitors have set to $0.1$ or $1.0$~pF respectively.
Those two configurations have also been compared to a RIS made by a PEC surface.
Simulations campaign are on progress and next steps are: i) optimize the diodes configurations by implementing optimization algorithm; ii) by considering a set of plane waves that act as random source; iii) consider more capacitors configurations.
FDTD simulation is an useful and reliable tool to speed up the analysis and the design of RISs.

\section*{Acknowledgment}
This work has been supported by EU H2020 RISE-6G project under the grant number 101017011.
We acknowledge PRACE for awarding us access to Joliot-Curie KNL at GENCI@CEA, France.

\end{document}